\documentclass[12pt,preprint]{aastex}

\shorttitle{(Running Header: COSMOLOGICAL ``CONSTANT'' AND ``BIG BOUNCE'')}
\shortauthors{Hongya Liu and Paul S. Wesson}

\begin{document}


\title{UNIVERSE MODELS WITH A VARIABLE COSMOLOGICAL ``CONSTANT''
 AND A ``BIG BOUNCE''}


\author{Hongya Liu \altaffilmark{1}}
\affil{Department of Physics, Dalian University of Technology\\
Dalian 116024, P.R. China}
\email{hyliu@dlut.edu.cn}

\author{Paul S. Wesson\altaffilmark{2}}
\affil{Department of Physics, University of Waterloo\\
Waterloo, Ontario N2L 3G1,Canada }
\email {wesson@astro.uwaterloo.ca}

\altaffiltext{1}{Corresponding author.}
\altaffiltext{2}{Fax: (519) 746 8115.}


\begin{abstract}
We present a rich class of exact solutions which contains
radiation-dominated and matter-dominated models for the early and late
universe. They include a variable cosmological ``constant'' which is derived
from a higher dimension and manifests itself in spacetime as an energy
density for the vacuum. This is in agreement with observational data and is
compatible with extensions of general relativity to string and membrane
theory. Our solutions are also typified by a non-singular ``big bounce'' (as
opposed to a singular big bang), where matter is created as in inflationary
cosmology.
\end{abstract}


\section{INTRODUCTION}

In general relativity, the cosmological constant $\Lambda $ may be regarded
as a measure of the energy density of the vacuum, and can in principle lead
to the avoidance of the big-bang singularity which is characteristic of
other Friedmann-Robertson-Walker (FRW) models. However, the rather
simplistic properties of the vacuum that follow from the usual form of
Einstein's equations can be made more realistic if that theory is extended,
which in general leads to a variable $\Lambda $. Recently, Overduin (1999)
has given an account of variable-$\Lambda $ models which have a non-singular
origin. Here we wish to approach this subject from a different perspective,
and give a class of exact solutions wherein $\Lambda $ can change in a
manner that is in agreement with observation but where the singular big bang
is replaced by a non-singular ``big bounce''.

There are several motives for this. In Einstein's theory, the density and
pressure of the vacuum are given in terms of the gravitational constant and
the speed of light by $\rho =-p/c^2=\Lambda /8\pi G$ (see e.g. Adler, Bazin
and Schiffer 1975 or Wesson 1999). However, astrophysical data constrain the
values of these to be many orders smaller than the corresponding quantities
inferred from particle physics, leading to the cosmological constant problem
(Weinberg 1989; Ng 1992; Adler, Casey and Jacob 1995). This can in principle
be resolved by introducing a variable scalar field to the right-hand side of
Einstein's equations. There are constraints on this approach, notably from
galaxy number counts, gravitational lens statistics and cosmic microwave
background anisotropies (for references see Overduin 1999; Overduin and
Cooperstock 1998). Also, while it is possible to modify 4D relativity by
introducing a scalar field, the latter is most naturally a part of $N(>4)$D
theory. For example, in 5D Kaluza-Klein theory the 10 potentials for the
gravitational field are augmented by the 4 potentials of electromagnetism
and a scalar potential which is related to mass (See Wesson 1999: even if
the electromagnetic potentials are absent and the scalar potential is a
constant, the 5D theory implies significant differences from the 4D theory).
Further, the horizon problem posed by the isotropy of the microwave
background in the context of the standard FRW models can be resolved by a
period of inflation in 4D which is most naturally driven by the dynamics of
higher dimensions (Linde 1990, 1991). Indeed, the most promising route to a
unification of classical gravity and electromagnetism with the quantum
theory of particle interactions is via $N$D manifolds, as in superstrings,
supergravity and membrane theory (West 1986; Green, Schwarz and Witten 1987;
Youm 2000). The latter theories, however, have yet to be developed to the
stage where they can be meaningfully tested. By comparison, the 5D theory is
a modest extension of 4D general relativity and is known to be in agreement
with the classical tests in the solar system (Kalligas, Wesson and Everitt
1995; Liu and Overduin 2000), as well as the less exact data from cosmology
(Wesson 1992). We will therefore in what follows go with the minimal
extension of general relativity, using previous work on 5D manifolds (Liu
and Mashhoon 1995, 2000) to see what effects an extra dimension has on $%
\Lambda $ and the big bang.

The plan of this paper is as follows. In Section 2 we will give a broad
class of 5D solutions and derive their associated 4D properties of matter.
In Section 3 we will focus for practical reasons on the sub-class of
solutions which is flat in 3D; and then in Sections 4 and 5 study this when
the equation of state is that of dust and radiation, respectively. Section 6
is a conclusion.

\section{5D SOLUTIONS AND THEIR 4D MATTER}

We choose units such that $8\pi G=c=1$, and let upper-case Latin indices run
0-4 and lower-case Greek indices run 0-3. Our coordinates are ($t,r,\theta,
\phi ,y$) with $d\Omega ^2\equiv \left( d\theta ^2+\sin ^2\theta d\phi
^2\right) $. The 5D line element $dS^2=g_{AB}dx^Adx^B$ contains the 4D one $%
ds^2=g_{\alpha \beta }dx^\alpha dx^\beta $. The \ 5D field equations in
terms of the Ricci tensor are $R_{AB}=0$. These contain the 4D field
equations, which in terms of the Einstein tensor and an induced
energy-momentum tensor are $G_{\alpha \beta }=T_{\alpha \beta }$. The
procedure to go from 5D to 4D is by now well known (see Liu and Mashhoon
1995 or Wesson 1999), and is guaranteed by Campbell's theorem. There are
many classes of solutions of the 5D field equations known, including the
much-discussed Ponce de Leon (1988) cosmologies.

A new class, which extends the FRW solutions and may be verified by
computer, is given by

\begin{eqnarray}
dS^2 &=&B^2dt^2-A^2\left( \frac{dr^2}{1-kr^2}+r^2d\Omega ^2\right) -dy^2 
\nonumber \\
A^2 &=&\left( \mu ^2+k\right) y^2+2\nu y+\frac{\nu ^2+K}{\mu ^2+k}  \nonumber
\\
B &=&\frac 1\mu \frac{\partial A}{\partial t}\equiv \frac{\stackrel{\bf %
\circ }{A}}\mu \hspace{1.5in}\quad \;\;\;.
\end{eqnarray}
Here $\mu =\mu (t)$ and $\nu =\nu (t)$ are arbitrary functions, $k$ is the
3D curvature index ($k=\pm 1,0$) and $K$ is a constant. After a lengthy
calculation we find that the 5D Kretschmann invariant takes the form 
\begin{equation}
I=R_{ABCD}R^{ABCD}=\frac{72K^2}{A^8}\;,
\end{equation}
which shows that $K$ determines the curvature of the 5D manifold (1). Note
that the result (2) agrees, after rescaling parameters, with that obtained
in the so-called canonical coordinates (Liu and Mashhoon 1995). From (1) we
can see that the form of $Bdt$ is invariant under an arbitrary
transformation $t=t(\widetilde{t})$. This gives us a freedom to fix one of
the two arbitrary functions $\mu (t)$ and{\bf \ }$\nu (t)$ without changing
the form of solutions (1). The other can be seen to relate to the 4D
properties of matter, to which we now proceed.

The 4D line element is 
\begin{equation}
ds^2=g_{\alpha \beta }dx^\alpha dx^\beta =B^2dt^2-A^2\left( \frac{dr^2}{%
1-kr^2}+r^2d\Omega ^2\right) .
\end{equation}
This has the Robertson-Walker form which underlies the standard FRW models,
and allows us to calculate the non-vanishing components of the 4D Ricci
tensor: 
\begin{eqnarray}
^{(4)}R_0^0 &=&-\frac 3{B^2}\left( \frac{\stackrel{\circ \circ }{A}}A-\frac{%
\stackrel{\circ }{A}\stackrel{\circ }{B}}{AB}\right)  \nonumber \\
^{(4)}R_1^1 &=&^{(4)}R_2^2=^{(4)}R_3^3=-\frac 1{B^2}\left[ \frac{\stackrel{%
\circ \circ }{A}}A+\frac{\stackrel{\circ }{A}}A\left( \frac{\stackrel{\circ 
}{2A}}A-\frac{\stackrel{\circ }{B}}B\right) +2k\frac{B^2}{A^2}\right] \;.
\end{eqnarray}
Now from (1) we have 
\[
B=\frac{\stackrel{\bf \circ }{A}}\mu \;,\qquad \stackrel{\circ }{B}=\frac{%
\stackrel{\bf \circ \circ }{A}}\mu -\frac{\stackrel{\bf \circ }{A}}\mu \frac{%
\stackrel{\bf \circ }{\mu }}\mu \quad . 
\]
Using these in (4), we can eliminate $B$ and $\stackrel{\circ }{B}$ from
them to give 
\begin{eqnarray}
^{(4)}R_0^0 &=&-\frac{3\mu \stackrel{\bf \circ }{\mu }}{A\stackrel{\circ }{A}%
}  \nonumber \\
^{(4)}R_1^1 &=&\,^{(4)}R_2^2=\,^{(4)}R_3^3=-\left[ \frac{\mu \stackrel{\bf %
\circ }{\mu }}{A\stackrel{\circ }{A}}+\frac{2\left( \mu ^2+k\right) }{A^2}%
\right] \,.
\end{eqnarray}
These yield the 4D Ricci scalar 
\begin{equation}
^{(4)}R=-6\left( \frac{\mu \stackrel{\bf \circ }{\mu }}{A\stackrel{\circ }{A}%
}+\frac{\mu ^2+k}{A^2}\right) \,.
\end{equation}
This together with (5) enables us to form the 4D Einstein tensor $%
^{(4)}G_\beta ^\alpha $ $\equiv {}^{(4)}\!R_\beta ^\alpha -\delta _\beta
^\alpha {}\,^{(4)}\!R/2$. Its non-vanishing components are 
\begin{eqnarray}
^{(4)}G_0^0 &=&\frac{3\left( \mu ^2+k\right) }{A^2}  \nonumber \\
^{(4)}G_1^1 &=&\,^{(4)}G_2^2=\,^{(4)}G_3^3=\frac{2\mu \stackrel{\bf \circ }{%
\mu }}{A\stackrel{\circ }{A}}+\frac{\mu ^2+k}{A^2}\quad .
\end{eqnarray}
These give the components of the induced energy-momentum tensor since
Einstein's equations $G_\beta ^\alpha =T_\beta ^\alpha $ hold.

Let us suppose that the induced matter is a perfect fluid with density $\rho 
$ and pressure $p$ moving with a 4-velocity $u^\alpha \equiv dx^\alpha /ds$,
plus a cosmological term whose nature is to be determined: 
\begin{equation}
^{(4)}G_{\alpha \beta }=(\rho +p)u_\alpha u_\beta +(\Lambda -p)g_{\alpha
\beta }\quad .
\end{equation}
As in the FRW models, we can take the matter to be comoving in 3D, so $%
u^\alpha =\left( u^0,0,0,0\right) $ and $u^0u_0=1$. Then (8) and (7) yield 
\begin{eqnarray}
\rho +\Lambda &=&\frac{3\left( \mu ^2+k\right) }{A^2}  \nonumber \\
p-\Lambda &=&-\frac{2\mu \stackrel{\bf \circ }{\mu }}{A\stackrel{\circ }{A}}-%
\frac{\mu ^2+k}{A^2}\quad .
\end{eqnarray}
These are the analogs for our solution (1) of the Friedmann equations for
the FRW solutions. As there, we are free to choose an equation of state,
which we take to be the isothermal one 
\begin{equation}
p=\gamma \rho \quad .
\end{equation}
Here $\gamma $ is a constant, which for ordinary matter lies in the range
(dust) $0\leq \gamma \leq 1/3$ (radiation or ultrarelativistic particles).
Using (10) in (9) we can isolate the density of matter and the cosmological
term: 
\begin{eqnarray}
\rho &=&\frac 2{1+\gamma }\left( \frac{\mu ^2+k}{A^2}-\frac{\mu \stackrel%
{\bf \circ }{\mu }}{A\stackrel{\circ }{A}}\right)  \nonumber \\
\Lambda &=&\frac 2{1+\gamma }\left[ \left( \frac{1+3\gamma }2\right) \left( 
\frac{\mu ^2+k}{A^2}\right) +\frac{\mu \stackrel{\bf \circ }{\mu }}{A%
\stackrel{\circ }{A}}\right] \;.
\end{eqnarray}
In these relations, $\mu =\mu (t)$ is still arbitrary and $A=A(t,y)$ is
given by (1). The matter density therefore has a wide range of forms and the
cosmological ``constant'' is variable.

Let us now consider singularities of the 5D manifold (1). Since metric (1)
is 5D Ricci flat, we have $R=0$ and $R^{AB}R_{AB}=0$. The third 5D invariant
is given in (2), from which we see that $A=0$ (with $K\neq 0$) corresponds
to a 5D singularity. This is a physical singularity and, as is in general
relativity, can be naturally explained as a ``big bang'' singularity. From
(6) and (11) we also see that if $\stackrel{\circ }{A}=0$ then all $^{(4)}R$%
, $\rho $ and $\Lambda $ become infinity. This means that $\stackrel{\circ }{%
A}=0$ (with $A,\mu ,\stackrel{\bf \circ }{\mu }\neq 0$) represents a second
kind of singularities where the 4D scalar curvature $^{(4)}R$, the 4D
induced mass density $\rho $ and $\Lambda $ diverge. Clearly, this is a 4D
singularity at which the 5D invariant (2) may keep normal. We can naturally
explain this one as a ``big bounce'' singularity at which the scale factor $A
$ reaches it's minimum with $\stackrel{\circ }{A}=0$. Note that $\stackrel{%
\circ }{A}=0$ (with $\mu \neq 0)$ implies $B=0$ by (1). So we conclude that 
{\it our 5D models (1) may have two kinds of singularities, ``big bang'' and
``big bounce'', characterized by }$A=0${\it \ and }$B=0${\it \ , respectively%
}. Here $A$ is the scale factor of the 3D space and we can call $B$ the
scale factor of the time. In the following sections we will see a
realization of the 5D ``big bounce'' models.

\section{THE SPATIALLY-FLAT MODEL}

Astrophysical data, including the age of the universe, are compatible with
an FRW model with flat space sections (Leonard and Lake 1995; Overduin
1999). We therefore put $k=0$ in (1), which gives 
\begin{eqnarray}
dS^2 &=&B^2dt^2-A^2\left( dr^2+r^2d\Omega ^2\right) -dy^2  \nonumber \\
A^2 &=&\left( \mu y+\frac \nu \mu \right) ^2+\frac K{\mu ^2}  \nonumber \\
B &=&\frac{\stackrel{\bf \circ }{A}}\mu \qquad \hspace{0.6in}\hspace{1in}%
\,\,\,\,.
\end{eqnarray}
The corresponding relations (11) become 
\begin{eqnarray}
\rho &=&\left( \frac 2{1+\gamma }\right) \frac \mu A\left( \frac \mu A-\frac{%
\stackrel{\bf \circ }{\mu }}{\stackrel{\circ }{A}}\right)  \nonumber \\
\Lambda &=&\left( \frac 2{1+\gamma }\right) \frac \mu A\left[ \left( \frac{%
1+3\gamma }2\right) \frac \mu A+\frac{\stackrel{\bf \circ }{\mu }}{\stackrel{%
\circ }{A}}\right] \;.
\end{eqnarray}
But as mentioned above, one of the functions $\mu (t)$ and $\nu (t)$ which
appear in (12) and (13) is free to choose without loss of generality. A
convenient choice and the corresponding form of $A(t,y)$ are given by: 
\begin{eqnarray}
\nu &=&-C\mu ^2t  \nonumber \\
A^2 &=&\mu ^2\left( y-Ct\right) ^2+\frac K{\mu ^2}\quad .
\end{eqnarray}
Here $C$ is a constant and $\mu =\mu (t)$ is the only function left
undetermined.

It is well known in general relativity that the isolation of a cosmological
term from densities $\rho $ and $p$ is not unique: one can always get $%
\Lambda $ absorbed into $\rho $ and $p$. The only thing which is of physical
importance is the equation of state satisfied by $\rho $ and $p$. This has
things to do with the vacuum energy. It is expected that the vacuum energy
could be contributed by either $\Lambda $ or dark matter, or by both.
Without a better model for the dark matter, we do not know how to separate
them. So we assume that {\it all constituents of }$\Lambda ${\it \ which
evolve like an ordinary matter through }$p=\gamma \rho $ {\it have already
been absorbed into }$\rho ${\it \ and }$p$. Furthermore, we also assume that
the remaining $\Lambda $ is small compared with $\rho $ in the present
universe. Then, to fix $\mu (t),$ we will impose a condition on the
cosmological term $\Lambda $ in equations (13) that as the model expands to
infinity, $\Lambda $ tends to zero more rapidly than $\rho $. [The relative
sizes of $\Lambda $ and $\rho $ will then be given by (13) and depend on the
epoch.] To implement this condition, we recall that in the $k=0$ FRW model
the scale factor is of a power law. This reminds us to try to use a power
law for $\mu (t)$ in (14), which give. 
\begin{eqnarray}
\mu (t) &=&t^n  \nonumber \\
A^2 &=&t^{2n}\left( y-Ct\right) ^2+Kt^{-2n}\quad .
\end{eqnarray}
The last of these shows that the power $n$ determines the behavior of the
scale factor $A(t,y)$ on the hypersurface $y=$ constant which is spacetime.
Specifically, we find that $A\propto t^{n+1}$ for $n\geqslant -1/2$ as $%
t\rightarrow \infty $. This into (13) gives 
\begin{eqnarray}
\rho &\propto &\left( 1-\frac n{n+1}\right) \frac 1{t^2}\quad ,\;\;%
t\rightarrow \infty  \nonumber \\
\Lambda &\propto &\left( \frac{1+3\gamma }2+\frac n{n+1}\right) \frac
1{t^2}\quad ,\;\;t\rightarrow \infty \quad .
\end{eqnarray}
These are observationally acceptable behaviors for ordinary matter (Leonard
and Lake 1995) and the cosmological term (Overduin 1999). However, the
condition that $\Lambda $ decays faster than $\rho $ means that we must have 
$n=-(1+3\gamma )/3(1+\gamma )$. This with $A\propto t^{n+1}$ and $\mu =t^n$
in (12) gives $B^2\rightarrow $ constant , for\ $t\rightarrow \infty $. This
means that the usual proper time is recovered if we let $B\rightarrow 1$ for 
$t\rightarrow \infty $. Then the constant $C$ in (14) is 
\begin{equation}
C=\left\{ 
\begin{array}{l}
\frac 32(1+\gamma )\;\quad \;for \; 0\leq \gamma <\frac 13 \\ 
\sqrt{4-K}\;\quad \; for \; \gamma =\frac 13\quad \qquad .
\end{array}
\right.
\end{equation}
We see that the value of this parameter depends on whether the matter is
cold or hot.

To complete this section, it is useful to recap the analysis and state the
result explicitly: The class of exact solutions (1) is very rich
algebraically; so we appeal to astrophysics to argue that $k=0$ which gives
(12), and to argue that $\Lambda $ decays faster than $\rho $ which fixes $%
\mu (t)$ and gives (17). The result is a spatially-flat cosmological
solution with good physical properties given by 
\begin{eqnarray}
dS^2 &=&B^2dt^2-A^2\left( dr^2+r^2d\Omega ^2\right) -dy^2  \nonumber \\
A^2 &=&\mu ^2\left( y-Ct\right) ^2+\frac K{\mu ^2}  \nonumber \\
B &=&\frac{\stackrel{\bf \circ }{A}}\mu  \nonumber \\
\mu (t) &=&t^n\;,\quad \;n=-\frac{1+3\gamma }{3(1+\gamma )}  \nonumber \\
C &=&\left\{ 
\begin{array}{l}
\frac 32(1+\gamma )\quad \; for \;0 \leq \gamma <\frac 13 \\ 
\sqrt{4-K}\quad \;for \; \gamma =\frac 13
\end{array}
\right. \quad \quad .
\end{eqnarray}
This solution has some features in common with others in the literature and
some which are different. A catalog of exact solutions of the 5D field
equations ($R_{AB}=0$) has been given by Wesson (1999), who has also
discussed the standard technique wherein these equations are reduced to
Einstein's ($G_{\alpha \beta }=T_{\alpha \beta }$) to obtain the properties
of matter. In this regard, it is important to note that matter is related to
the 4D space and its curvature. Thus the much-discussed 5D cosmologies due
to Ponce de Leon (1988) are flat in 5D but curved in 4D. That is why they
contain matter. However, the Ponce de Leon cosmologies are separable in $t$
and $y$, reducing to FRW ones on the hypersurfaces $y=$ constants. In
contrast, the cosmologies (1) and their 3D-flat subset (18) are not
separable. To distinguish between models of the Ponce de Leon kind and those
discussed here will require new methods which probe the dynamics associated
with the $y$-dependence of the solutions. This could involve new tests of
gravity in the solar system, such as the planned Space Test of the
Equivalence Principle (Wesson 1996; Overduin 2000). But tests of 5D models
could also be made astrophysically, and to this end we will examine in the
following two sections the implications of (18) for cold and hot matter.

\section{THE COLD 3D-FLAT MODEL}

For $\gamma =0$ in (10) the equation of state is $p=0$ for dust, which is
believed to describe the matter content of the late universe. In this case
by (18) we have $n=-1/3$, $C=3/2$ for the constants and $\mu (t)=t^{-1/3}$
for the function which describes the scale factor. The latter is given
explicitly by

\begin{equation}
A^2=\frac 94t^{4/3}+Kt^{2/3}-3yt^{1/3}+y^2t^{-2/3}\quad .
\end{equation}
We have examined this numerically for various values of the hypersurfaces $%
y= $ constants and the three possible values of the normalized 5D curvature
parameter $K$ which appears in (2), namely $K=\pm 1,0$. Note that $y=0$ is
too special to represent a typical $y=$ constant hypersurface, because in
that case the last two terms in (19) do not contribute to the scale factor $%
A $. We find that in most other cases the results are similar. Here we
present only one for the purpose of illustration. Thus Figure $1$ is a plot
of $A$ versus $t$ for $K=1$ and $y=-1,-2/3,-1/3$, respectively.
We see that there is a finite minimum of $A$ at $t=t_m$, before which the
universe contracts and after which it expands. This kind of behavior can
happen in 4D FRW models which have a large value of a constant $\Lambda $
(Robertson and Noonan 1968). But most FRW models have a singularity in the
geometry and divergent properties of matter because $A(t)\rightarrow 0$ for $%
t\rightarrow 0$, defining the big bang. Here the situation is different. By
(2), the 5D geometry would be singular if $A(t)\rightarrow 0$, but by (19)
this would only happen on the special hypersurfaces defined by the roots of
that equation. In general, on $y=$ constant hypersurfaces the 5D curvature
invariant (2) is finite. However, at $t=t_m$ we have $\stackrel{\bf \circ }{A%
}=0$ and $\mu (t)=t^{-1/3}\neq 0$. So by (18) we have $B=0$ at $t=t_m$.
Thus, as discussed in section 2, we get a ``big bounce'' model in which all $%
^{(4)}R$, $\rho $ and $\Lambda $ diverge at $t=t_m$ except the 5D invariant
(2), which keeps normal.

To investigate this and other aspects of the model, we note from Figure 1
that $t_m$ is of order unity so we can study the cases $t\gg t_m$ and $t\ll
t_m$.

For $t\gg t_m$ (19) gives 
\begin{equation}
A=\frac 32t^{2/3}\left[ 1+\frac 29Kt^{-2/3}-\frac
23yt^{-1}+O(t^{-4/3})\right] \;,
\end{equation}
whose derivative in (18) gives 
\begin{equation}
B=\frac{\stackrel{\bf \circ }{A}}\mu =1+\frac y{3t}+O(t^{-4/3})\;.
\end{equation}
The 5D line element of (18) now reads 
\begin{equation}
dS^2=\left[ 1+O(t^{-1})\right] dt^2-\left[ \frac 32t^{2/3}+\frac
13K+O(t^{-1/3})\right] ^2\left( dr^2+r^2d\Omega ^2\right) -dy^2\;.
\end{equation}
This is similar to the 4D Einstein-de Sitter metric, insofar as $t$ tends to
the proper time and the scale factor then varies as $t^{2/3}$. Using the
constants noted at the beginning of this section and $A$ of (20), we can
obtain the density of matter and the cosmological term from (13). They are 
\begin{eqnarray}
\rho &=&\frac 4{3t^2}\left[ 1-\frac{10}{27}Kt^{-2/3}+O(t^{-1})\right] 
\nonumber \\
\Lambda &=&-\frac{8K}{81t^{8/3}}\left[ 1+O(t^{-1/3})\right] \;\qquad \qquad .
\end{eqnarray}
From this and (20) we find that 
\begin{eqnarray}
\rho A^3 &=&\frac 92\left[ 1+\frac 8{27}Kt^{-2/3}+O(t^{-1})\right]  \nonumber
\\
\frac \partial {\partial t}\left( \rho A^3\right) &=&-\frac
89Kt^{-5/3}\left[ 1+O(t^{-1/3})\right] \;\qquad .
\end{eqnarray}
This means that the mass of the (uniform) cosmological fluid within a
spherical volume defined by the scale factor changes with time. By
comparison with the corresponding situation in general relativity, this may
be thought of as being due to the effective pressure associated with the
cosmological term (Henriksen, Emslie and Wesson 1983). By (23) and (24), for 
$K>0$ we have $\Lambda <0$ and $\left( \rho A^3\right) ^{\circ }<0$, while
for $K<0$ we have $\Lambda >0$ and $\left( \rho A^3\right) ^{\circ }>0$. [If 
$K=0$, (23) has to be calculated to higher order.] The generation of mass in
quantum field theory has been studied before, particularly in the context of
inflationary cosmology (Linde 1990). Here we have a classical analog of that
process due to a variable cosmological ``constant'', that operates even at
late times.

For $t\ll t_m$ (19) gives 
\begin{equation}
A=yt^{-1/3}\left[ 1-\frac 3{2y}t+\frac K{2y^2}t^{4/3}+O\left( t^2\right)
\right] \;,
\end{equation}
whose derivative in (18) gives 
\begin{equation}
B=\frac{\stackrel{\bf \circ }{A}}\mu =-\frac y{3t}\left[ 1+\frac 3yt-\frac{3K%
}{2y^2}t^{4/3}+O\left( t^2\right) \right] \;.
\end{equation}
The 5D line element of (18), correct to the leading order in (25), now reads 
\begin{equation}
dS^2=\left( \frac y{3t}\right) ^2dt^2-\left( yt^{-1/3}\right) ^2\left(
dr^2+r^2d\Omega ^2\right) -dy^2\quad .
\end{equation}
If we carry out the coordinate transformation $t=L^3e^{3\tau /L}$ on this
(where $L$ is a constant), it becomes 
\begin{equation}
dS^2=\frac{y^2}{L^2}\left[ d\tau ^2-e^{-2\tau /L}\left( dr^2+r^2d\Omega
^2\right) \right] -dy^2\quad .
\end{equation}
This is a metric of the canonical form (Liu and Mashhoon 1995; Wesson 1999).
And the part inside the square brackets is in fact the de Sitter metric,
which in 4D would be interpreted as having $\rho =0$, $\Lambda =3/L^3$.
Here, however, the density and cosmological term are obtained as before by
substituting (25) and the other appropriate quantities into (13). They are 
\begin{eqnarray}
\rho &=&\frac 9{y^3}t\left[ 1-\frac{4K}{9y}t^{1/3}+O\left( t\right) \right] 
\nonumber \\
\Lambda &=&\frac 3{y^2}\left[ 1+\frac K{3y^2}t^{4/3}+O\left( t^2\right)
\right] \;.
\end{eqnarray}
These show that $\rho \rightarrow 0$ and $\Lambda \rightarrow 3/y^2$ for $%
t\rightarrow 0$, so the universe was empty of matter and $\Lambda $%
-dominated at its start in $t$-time. However, (28) shows that the proper
time in 4D is not $t$ but $\tau $ , where by our coordinate transformation $%
t=0$ corresponds to $\tau =-\infty $. So in $\tau $-time the universe was
empty and existed forever, before its matter was produced around the ``big
bounce''. The generation of matter by quantum tunneling has been studied
before, as a means of creating universes from nothing (Vilenkin 1982). Here
we have a classical analog of that process.

\section{THE HOT 3D-FLAT MODEL}

For $\gamma =1/3$ in (10) the equation of state is $p=\rho /3$ for radiation
or ultrarelativistic particles, which is believed to describe the matter
content of the early universe. In this case by (18) we have $n=-1/2$, $%
C=\left( 4-K\right) ^{1/2}$ for the constants and $\mu (t)=t^{-1/2}$ for the
function that describes the scale factor. The latter is given explicitly by 
\begin{equation}
A^2=t^{-1}(y-Ct)^2+Kt=4t-2Cy+\frac{y^2}t\;.
\end{equation}
Figure 2 is a plot of $A$ versus $t$ for $K=1$ and $y=-1,-2/3,-1/3$,
respectively.
We see that the behavior is similar to that for the $\gamma =0$ model
studied in the preceding section, with a ``big bounce'' at $t=t_m$.

For $t\gg t_m$ (30) gives 
\begin{equation}
A=2t^{1/2}\left[ 1-\frac{Cy}{4t}+O(t^{-2})\right] \;,
\end{equation}
whose derivative in (18) gives 
\begin{equation}
B=\frac{\stackrel{\bf \circ }{A}}\mu =1+\frac{Cy}{4t}+O(t^{-2})\;.
\end{equation}
The 5D line element of (18) now reads 
\begin{equation}
dS^2=\left[ 1+\frac{Cy}{2t}+O(t^{-2})\right] dt^2-4t\left[ 1-\frac{Cy}{2t}%
+O(t^{-2})\right] \left( dr^2+r^2d\Omega ^2\right) -dy^2\;.
\end{equation}
This is similar to the 4D radiation metric, insofar as $t$ tends to the
proper time and the scale factor then varies as $t^{1/2}$. We also note that
the combination $y/t$ in (33) is self-similar, which is a characteristic of
4D astrophysical systems (Henriksen, Emslie and Wesson 1983). Using the
constants noted at the beginning of this section and $A$ of (31), we can
obtain the density of matter and the cosmological term from (13). They are 
\begin{eqnarray}
\rho &=&\frac 3{4t^2}\left[ 1+\frac{Cy}{4t}+O(t^{-2})\right]  \nonumber \\
\Lambda &=&\frac{3Cy}{16t^3}\left[ 1+O(t^{-1})\right] \qquad \quad .
\end{eqnarray}
>From this and (31) we find that 
\begin{equation}
\rho A^4=12\left[ 1-\frac{3Cy}{4t}+O(t^{-2})\right] \;.
\end{equation}
This is similar to the situation in general relativity, where $\rho A^4=$
constant because the number density of photons decreases as $A^3$ due to the
expansion and their energy decreases as $A$ due to the redshift.

For $t\ll t_m$ (30) gives 
\begin{equation}
A=yt^{-1/2}\left[ 1-\frac Cyt+O(t^2)\right] \;,
\end{equation}
whose derivative in (18) gives 
\begin{equation}
B=\frac{\stackrel{\bf \circ }{A}}\mu =-\frac y{2t}\left[ 1+\frac
Cyt+O(t^2)\right] \;\quad .
\end{equation}
The 5D line element of (18), correct to the leading order in (36), now reads 
\begin{equation}
dS^2=\left( \frac y{2t}\right) ^2dt^2-y^2t^{-1}\left( dr^2+r^2d\Omega
^2\right) -dy^2\;.
\end{equation}
If we carry out the coordinate transformation $t={\cal L}^2e^{2\tau /{\cal L}%
}$ on this (where ${\cal L}$ is a constant), it becomes 
\begin{equation}
dS^2=\frac{y^2}{{\cal L}^2}\left[ d\tau ^2-e^{-2\tau /{\cal L}}\left(
dr^2+r^2d\Omega ^2\right) \right] -dy^2\quad .
\end{equation}
This is a metric of the canonical form, and in fact the same as (28). The
density and cosmological term are again obtained by substituting (36) and
the other appropriate quantities into (13). They are 
\begin{eqnarray}
\rho &=&\frac{3C}{y^3}t\left[ 1+O(t)\right]  \nonumber \\
\Lambda &=&\frac 3{y^2}\left[ 1+\frac Cyt+O(t^2)\right] \;.
\end{eqnarray}
These show that $\rho \rightarrow 0$ and $\Lambda \rightarrow 3/y^2$ for $%
t\rightarrow 0$, which is the same as in the $\gamma =0$ case considered in
the preceding section. [However, (29) and (40) are different for finite $t$%
..] Similar comments apply here as there concerning the generation of matter
around the ``big bounce''.

\section{CONCLUSION}

We have given a class (1) of exact solutions of the 5D field equations which
extends the class of 4D Friedmann-Robertson-Walker solutions. These
solutions, unlike others in the literature including the Ponce de Leon
cosmologies, are not separable. But as with other 5D solutions, their matter
content can be obtained by a standard technique, which gives the density,
pressure and cosmological ``constant'' (9). The last, however, is really a
measure of the variable energy density of the vacuum. This generalizes the
vacuum of general relativity, and is in conformity with other
higher-dimensional extensions of that theory such as strings and membranes.
The class of solutions we have discussed is algebraically rich, so we have
used input from cosmology to study its properties. The subclass with
spatially-flat geometry (12) yields a cold model (19) which is the analog of
the 4D dust one, and a hot model (30) which is the analog of the 4D one for
radiation or ultrarelativistic particles. Both models have a cosmological
``constant'' that decays faster than the density decreases, in accordance
with astrophysical data including the age of the universe. Note that our
cosmological ``constant'' do not include those parts of the conventional $%
\Lambda $ which evolve like an ordinary matter and can be absorbed into $%
\rho $ and $p$. Also, both models in general show a big ``bounce'' as
opposed to a big bang, where the 5D geometry is non-singular but where the
matter properties and the 4D scalar curvature diverge. We therefore recover
the physics of the late and early universe and the fireball, but without a
breakdown in the geometry. In view of the major implications of a bounce as
opposed to a bang, we have examined the behavior of the models around that
event in some detail. The main feature is that matter is created during the
bounce, in agreement with models of inflation based on quantum field theory.

It is known that a major difficulty of the 4D big bounce models in standard
general relativity (Robertson and Noonan 1968) is that the collapsing phase
would typically generate enormous inhomogeneities. Here we wish to emphasize
that our 5D big bounce models are different in many aspects from the
conventional 4D big bounce models. Firstly, our models are not symmetric
before and after the big bounce. Secondly, in our models, the contraction
and expansion of the universe around the bounce was accompanied by matter
creation. If the process of the contraction of the universe before the
bounce was dominated not by collapsing of previously formed matter but by
creating of new matter, (this is very possible because in our models the
universe was empty at the beginning of the contracting,) then hopefully the
inhomogeneity problem could be resolved.

We are aware that the present work is exploratory. While our solutions are
quite general, we have restricted their application by assuming that
ordinary matter and the vacuum can be described by a perfect fluid, by
adopting an isothermal equation of state for the matter, by concentrating on
the spatially flat case, and that only for dust and radiation. All of these
restrictions could be removed in future work. However, we have shown that
adding only one extra dimension to general relativity makes the big bang a
subject of fruitful analysis and not just a no-go singularity.

\bigskip

We thank the referee and editor for helpful comments. This work grew out of
a previous collaboration with B. Mashhoon and was supported financially by
NSF of P.R. China and NSERC of Canada.

\bigskip
\clearpage


\begin{figure}[tbp]
\plotone{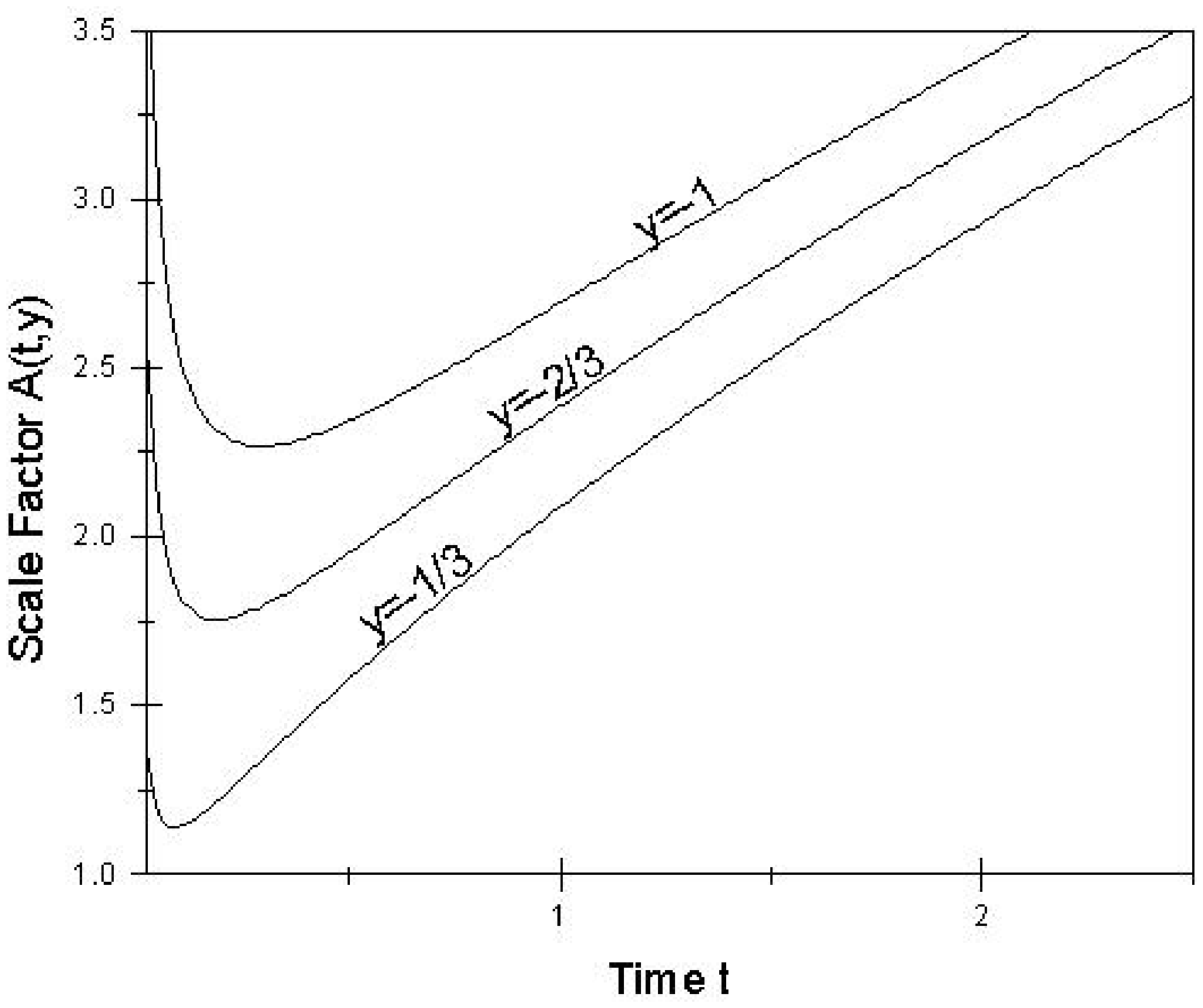}
\caption{Evolution of the scale factor $A$ for the \textbf{cold} 3D-flat
model with $K=1$ and $y=-1,-2/3,-1/3$.}
\label{fig1}
\end{figure}

\clearpage 

\begin{figure}[tbp]
\plotone{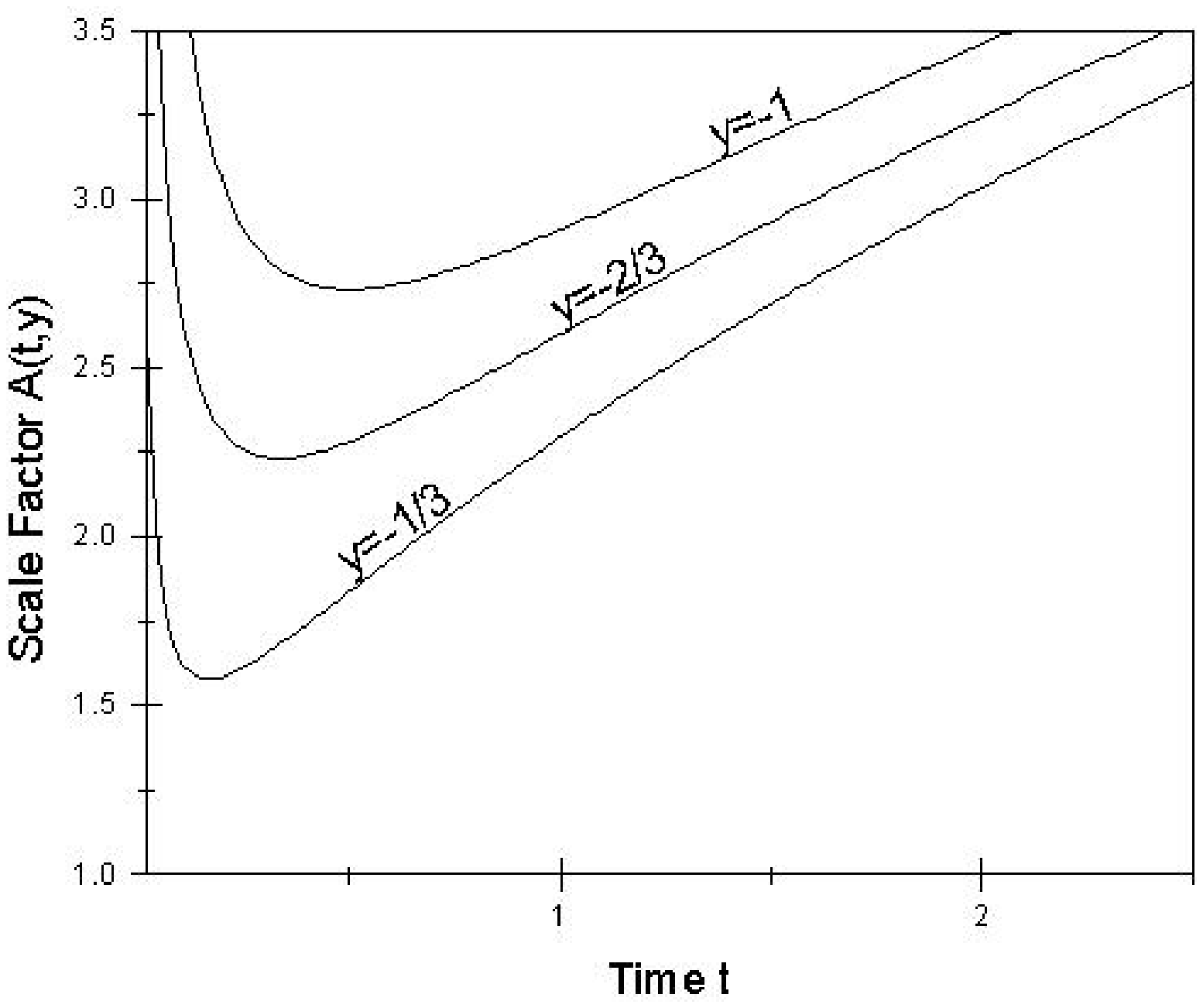}
\caption{ Evolution of the scale factor $A$ for the \textbf{hot} 3D-flat
model with $K=1$ and $y=-1,-2/3,-1/3$.}
\label{fig2}
\end{figure}

\end{document}